\documentclass[12pt]{article}
\usepackage{a4wide,amssymb,cite}
\pdfoutput=1 

\usepackage[T1]{fontenc} 

\usepackage[symbol*]{footmisc}
\usepackage{amssymb}
\usepackage{epsfig} 
\usepackage{amsmath} 
\usepackage{graphicx}

\usepackage{a4wide,amssymb,cite,graphicx}
\usepackage{epsfig}
\usepackage[usenames,dvipsnames]{color}
\usepackage{slashed}
\usepackage{ulem}

\newcommand{\beq}{\begin{equation}}
\newcommand{\eeq}{\end{equation}}
\newcommand{\bea}{\begin{eqnarray}}
\newcommand{\eea}{\end{eqnarray}}
\newcommand{\ba}{\begin{array}}
\newcommand{\ea}{\end{array}}
\newcommand{\bi}{\begin{itemize}}
\newcommand{\ei}{\end{itemize}}
\newcommand{\bn}{\begin{enumerate}}
\newcommand{\en}{\end{enumerate}}
\newcommand{\bc}{\begin{center}}
\newcommand{\ec}{\end{center}}
\renewcommand{\l}{\left}
\renewcommand{\r}{\right}

\newcommand{\eq}[1]{Eq.~(\ref{#1})}
\newcommand{\eqs}[2]{Eqs.~(\ref{#1}) and (\ref{#2})}
\newcommand{\eqss}[3]{Eqs.~(\ref{#1}), (\ref{#2}) and (\ref{#3})}

\newcommand{\eV}{\mathinner{\mathrm{eV}}}
\newcommand{\keV}{\mathinner{\mathrm{keV}}}
\newcommand{\MeV}{\mathinner{\mathrm{MeV}}}
\newcommand{\GeV}{\mathinner{\mathrm{GeV}}}

\newcommand{\pl}{{\rm Pl}}

\DefineFNsymbolsTM{myfnsymbols}{
  \textasteriskcentered *
  \textdagger    \dagger
  \textdaggerdbl \ddagger
  \textsection   \mathsection
  \textbardbl    \|%
  \textparagraph \mathparagraph
}%

\setfnsymbol{myfnsymbols}

\begin{document}

\begin{titlepage}

\rightline{KIAS-P14012}

\begin{centering}
\vspace{1cm}
{\large {\bf Cluster X-ray line at $3.5\,{\rm keV}$ from axion-like dark matter    }} \\

\vspace{1.5cm}

{\bf Hyun Min Lee}$^{1}$,  {\bf Seong Chan Park}$^{2}$, and  {\bf Wan-Il Park}$^{3}$   \footnote{All authors contributed equally.}\\
\vspace{.5cm}

{\it $^1$Department of Physics, Chung-Ang University, Seoul 156-756, Korea.} 
\\  \vspace{.5mm}
{\it $^2$Department of Physics, Sungkyunkwan  University, Suwon 440-746, Korea.}
\\  \vspace{.5mm}
{\it $^3$School of Physics, KIAS, Seoul 130-722, Korea.}  
\\

\vspace{.1in}

\end{centering}
\vspace{2cm}

\begin{abstract}
\noindent

The recently reported X-ray line signal at $E_\gamma \simeq 3.5\, {\rm keV}$ from a stacked spectrum of various galaxy clusters and the Andromeda galaxy may be originating from a decaying dark matter particle of the mass $2 E_\gamma$. A light axion-like scalar is suggested as a natural candidate for dark matter and its production mechanisms are closely examined. We show that the right amount of axion relic density with the preferred parameters, $m_a \simeq 7 \,{\rm keV}$ and $f_a \simeq 4\times 10^{14}\, {\rm GeV}$, can be naturally obtainable from the decay of inflaton.
If the axions were produced from the saxion decay, it could not have constituted the total relic density due to the bound from structure formation. 
Nonetheless, the saxion decay is an interesting possibility, because the $3.5\, {\rm keV}$ line and dark radiation can be addressed simultaneously, being consistent with the Planck data.  Small misalignment angles of the axion, ranging between $\theta_a\sim 10^{-4} -10^{-1}$ depending on the reheating temperature, can also be the source of axion production. The model with axion misalignment can satisfy the constraints for structure formation and iso-curvature perturbation.

\end{abstract}

\vspace{3cm}

\end{titlepage}

\section{Introduction}

It has been recently reported by two groups that there exists a line signal at 3.5 $\keV$ in a stacked X-ray spectrum of galaxy clusters and the Andromeda galaxy \cite{Xray1,Xray2}. 
Since no source of X-ray lines, such as atomic transitions, is known at this energy, the observed line may suggest the existence of a new source. It would be tantalizing to notice that the line is consistent with the monochromatic photon signal due to a decaying dark matter(DM) with the mass, $m_{\rm DM}\simeq 7\,{\rm keV}$, and the lifetime, $\tau_{\rm DM}\simeq10^{28}\,{\rm sec}$ \cite{Xray1,Xray2}. 
Even though a further confirmation of the line by independent and refined observations, e.g. Astro-H mission \cite{astroH}, is required, it would be timely and interesting to investigate the possible DM candidates with existing data. One obvious candidate is the sterile neutrino in models for the neutrino masses \cite{Xray1,Xray2,takahashi1} but there could be alternative DM explanations\cite{finkbeiner}.  

In this paper, we propose an axion-like particle as the source of the $X$-ray line at 3.5 $\keV$.\footnote{We note that there have appeared related papers on the keV axion dark matter \cite{takahashi2,ringwald} while we were finalizing our work.} Axion-like particles (or simply axions) are ubiquitous in various extensions of the SM including stringy models with various compactification schemes where pseudo-Goldstone bosons appear in the low energy after the breakdown of accidental global symmetries \cite{axionreview}.  
The mass spectra of the axion-like particles can span a wide range, covering the $\keV$ scale.
We may identify one of them as the axion-like particle explaining the X-ray line.\footnote{The axion-like particles, in general, do not play the role of the QCD axion.} 
The axion-like particle could be produced in the early universe by various mechanisms and account for the observed DM amount: e.g. the axion misalignment, the decay of the radial partner of the axion-like DM, dubbed the saxion, and also the decay of the inflaton. We show that an axion-like DM, produced preferably in the decay of the inflaton, can provide a good fit to the observed X-ray line by its decay into a pair of photons through anomaly interactions with the decay constant, $f_a\simeq 4\times 10^{14}\,{\rm GeV}$. 

The rest of the paper is organized as follows. In Sect.~\ref{sec:model}, we describe our model for an axion-like DM and discuss the cosmological bounds from structure formation and iso-curvature perturbation in Sect.~\ref{sec:cosmo}. In  Sect.~\ref{sec:prod}, three scenarios of axion production and relevant observational bounds on them are discussed.  Finally, conclusions are drawn in Sect.~\ref{sec:conclusion}.

\section{The model with a 7 $\keV$ axion for the $X$-ray line} \label{sec:model}

We introduce an axion-like particle as a pseudo-Goldstone boson associated with a broken anomalous $U(1)$ symmetry.
After the symmetry breaking, the model-independent effective Lagrangian for the axion $a$ and the saxion $s$ (the radial component of the symmetry-breaking field) can be expressed as
\bea \label{Lag}
{\cal L} 
&=& \frac{1}{2}(\partial_\mu s)^2 +\frac{1}{2}(\partial_\mu a)^2+ \frac{s}{2 f_a} (\partial_\mu a)^2 - \frac{1}{2} m^2_s s^2 - \frac{1}{2} m^2_a a^2 - \frac{1}{4} F_{\mu\nu} F^{\mu\nu}
\nonumber \\
&& + \frac{c_1 \alpha_{\rm em}}{8 \pi f_a} ( a F_{\mu\nu} \tilde{F}^{\mu\nu} - s F_{\mu\nu} F^{\mu\nu} ) +  \frac{c_2}{f_a} (\partial_\mu a) \bar{f} \gamma^\mu \gamma^5 f  + i  \bar{f} \slashed{D}_\mu \gamma^\mu f - m_f f \bar{f} 
\nonumber \\
&& -  (m_f e^{c_3 (s+i a) / f_a} \bar{f}_L f_R + h.c. ) + {\cal L}_I
\eea
where $\alpha_{\rm em}$ is the fine structure constant of electromagnetic interaction, $f_a$ is the axion coupling constant, and $F_{\mu\nu}$ and $\tilde{F}_{\mu\nu}$ are  the field-strength tensor and its dual for electromagnetic field, respectively.
We note that $c_{1,2,3}$ are dimensionless parameters of order one and we have included an extra charged fermion $f$ that is responsible for the generation of anomalies.
For example, in a  supersymmetric axion model \cite{susyaxion}, the axion chiral multiplet $A$ with $A|_{\theta=0}=s+ia$ appears in the superspace interactions as $\int d^2\theta A W^\alpha W_\alpha$ and $\int d^2\theta\, m_f \, e^{c_3 A/ f_a}\Phi \bar{\Phi}$ where $W_\alpha$ is the field strength superfield, and $\Phi$ and $\bar{\Phi}$ are matter chiral multiplets containing the extra charged fermion. But the results in our work do not depend on the presence of supersymmetry. 
We can add the Lagrangian for inflation, ${\cal L}_{I}$.

For a $\keV$-scale axion, the first term of the second line in \eq{Lag} provides the main decay channel with a rate given by
\begin{eqnarray}
\Gamma_{a\to \gamma\gamma} = \frac{\alpha_{\rm em}^2 m_a^3}{64 \pi^3 f_a^2}.
\end{eqnarray}
where we set $c_1=1$ for simplicity.
This axion decay can be a possible origin of the recently reported $X$-ray line at $3.5
\,{\rm keV}$,  if the axion saturates the dark matter relic density and has the following properties \cite{Xray1,Xray2}
\beq
m_a = 7.1 \keV, \quad \tau_a = 1.14 \times 10^{28} {\rm sec} \quad {\rm or}\quad \Gamma_a = 5.73 \times 10^{-53} \GeV
\eeq
where $\tau_a$ is the lifetime of the axion. 
This implies that the axion coupling constant should be 
\beq \label{fa}
f_a \simeq 4 \times 10^{14} \GeV \l( \frac{m_a}{7 \keV} \r)^{3/2}.
\eeq
For notational convenience in the later section, we use $f_{a,0} \equiv 4 \times 10^{14} \GeV$.

\section{Cosmological constraints} 
\label{sec:cosmo}
Our $\keV$-scale axion can be constrained by astrophysics and cosmology, namely, the structure formation or the iso-curvature perturbation of dark matter, depending on how the axion is produced.

\begin{itemize}
\item{Structure formation}

The $\keV$ axion may be a decay product of the inflaton and/or the saxion.
Suppose that a mother particle, denoted as $X$, decays to two axions, each of which carries the energy of
\beq \label{pa-ini}
E_{a, i} = m_X/2
\eeq
when the axion mass is ignored.
The axion of our interest is expected to be out of thermal equilibrium at temperatures well below GUT scale \cite{ringwald}.
Hence the momentum of the axion is simply red-shifted once it is produced from the decay of $X$.
Then, in order not to destruct large scale structures, the axion should be non-relativistic around the epoch when a Hubble patch contains energy corresponding to the galactic-sized halo (corresponding to $T \sim T_* \equiv 300 \eV$) (see for example \cite{WDMconstraint1,WDMconstraint2}).

More precisely, the most recent analysis of Lyman-$\alpha$ forest data shows that the comoving free-streaming length of DM at the matter-radiation equality is constrained to be at $95\%$ C. L. \cite{Viel:2013fqw}
\beq \label{lfs-bnd}
\lambda_{\rm fs} <  \lambda^c_{\rm fs} \approx 0.12\, {\rm Mpc} 
\eeq
where we introduced $\lambda_{\rm fs}^{c}$ representing the observational bound on the free-streaming length.
Including the previous various analyses of Lyman-$\alpha$ forest data \cite{lyalpha} and the phase space densities derived from the dwarf galaxies of the Milky way \cite{dSph}, leads to the bound on the free-streaming length ranging between $0.12\,{\rm Mpc}\lesssim \lambda^c_{\rm fs}\lesssim 0.60\,{\rm Mpc}$.
The comoving free-streaming length of the axion produced from the decay of a mother particle $X$ is  computed as
\bea \label{lfs} \label{lfs}
\lambda_{\rm fs} 
&\equiv& \int_{\tau_X}^{t_{\rm eq}} \frac{v_a dt}{R(t)} = \int_{\tau_X}^{t_{\rm nr}} \frac{dt}{R(t)} + \int_{t_{\rm nr}}^{t_{\rm eq}} \frac{v_a dt}{R(t)} 
\nonumber \\
&=& \frac{2 t_{\rm nr}}{R_{\rm nr}} \l[ 1 - \l( \frac{\tau_X}{t_{\rm nr}} \r)^{1/2} + \frac{1}{2} \ln \l( \frac{t_{\rm eq}}{t_{\rm nr}} \r) \r]
\nonumber \\
&\approx& \frac{1}{H_0} \l( \frac{H_0}{\Gamma_X} \r)^{1/2} \l( \frac{m_X/2}{m_a} \r) \l( \frac{T_{\rm eq}}{T_0} \r)^{1/4} \l\{ 1 + \frac{1}{2} \ln \l[ \frac{\Gamma_X}{H_0} \l( \frac{m_a}{m_X/2} \r)^2 \l( \frac{T_0}{T_{\rm eq}} \r)^{3/2} \r] \r\}
\eea
where $v_a$ is the velocity of the axion, $R(t)$ is the scale factor, $t_{\rm nr}$ is the time when the axion becomes non-relativistic, and $\tau_X(\Gamma_X)$ is the lifetime(decay rate) of $X$, related to the decay temperature $T_X$ by $\tau_X=(\pi^2 g_*/90)^{-1/2} M_P/T^2_X$.
In the second line, we used $R(t) = R_{\rm eq} (t/t_{\rm eq})^{1/2}$ for $t < t_{\rm eq}$, and $v_a(t) = \l(t_{\rm nr} / t \r)^{1/2}$ for $t_{\rm nr} \lesssim t \lesssim t_{\rm eq}$.

The constraint \eq{lfs-bnd} is now interpreted as
\beq \label{TX-struc-bnd}
\frac{T_X}{m_X} \gtrsim 0.2 \l( \frac{0.12 {\rm Mpc}}{\lambda_{\rm fs}^{c}} \r) \l( \frac{200}{g_*(T_X)} \r)^{1/4} \l( \frac{7 \keV}{m_a} \r)
\eeq
where  $\l\{ \cdots \r\}$ in the last line of \eq{lfs} was approximated to $\l\{ \cdots \r\} = 8.27$.

\item{Isocurvature perturbation}

Axions in our scenario can be produced by either the decay of inflaton/saxion or the axion misalignment.
In the case of saxion decay, the iso-curvature perturbation of the saxion is potentially problematic.
However, it could be suppressed since the saxion could have a Hubble scale mass during inflation.
So, we consider only the case of axion misalignment in our discussion.

The recent Planck data combined with WMAP polarization data leads to a constraint on the fraction of the iso-curvature perturbation by \cite{Planckdata},
\beq
\frac{\mathcal{P}_S}{\mathcal{P}_\mathcal{R}} < 0.041,
\eeq
at 95\% CL, where $\mathcal{P}_\mathcal{R} \simeq 2.2 \times 10^{-9}$ and $\mathcal{P}_S$ are the power spectra of curvature and iso-curvature perturbations, respectively.
The iso-curvature perturbation of the axion dark matter can be expressed as \cite{axionisocurvature}
\beq
\mathcal{P}_S = \l(  r\, \frac{\partial \ln \Omega_a}{\partial \theta_{\rm osc}} \frac{H_I}{2 \pi f_a^I} \r)^2
\eeq
where $r$ is the fractional contribution of axion DM to the observed relic density of DM, $\theta_{\rm osc}$ is the initial misalignment angle,\footnote{We assume the misalignment angle is constant until the commencement of axion coherent oscillation.} and $H_I$ is the expansion rate during inflation, and $f_a^I$ is the axion coupling constant during inflation. 
As described in the next section, for $\theta_{\rm osc} \ll 1$, $\Omega_a \propto \theta_{\rm osc}^2$, hence one finds
\beq \label{HI-bnd}
H_I \lesssim \frac{1.2 \times 10^6 \GeV}{r} \l( \frac{\theta_{\rm osc}}{10^{-4}} \r) \l( \frac{f_a^I}{f_a} \r)
\eeq
Note that $f_a^I$ can be much larger than $f_a$ at zero temperature.
In addition, $\theta_{\rm osc}$ can be $\mathcal{O}(1)$ if the axion relic density can be diluted by some amount due to, for example, a late-time entropy production.
\end{itemize}

\section{Scenarios of axion production}
\label{sec:prod}

Axion-like scalars can be produced from decays of heavy particles or coherent oscillations.
In this section, we discuss how we can obtain a right amount of the $\keV$-scale axion while satisfying various constraints given in the previous section.
In particular, we consider the axion production from the inflaton/saxion decay and the axion misalignment in both cases of high and low reheating temperatures after primordial inflation. 
In order to match the relic density of dark matter to the observed one, $\Omega_a = 0.268$ \cite{Planckdata}, we quote the necessary axion abundance at present as   
\beq \label{Ya-obs}
Y_a \simeq 6.9 \times 10^{-5} \l( \frac{7 \keV}{m_a} \r).
\eeq

\subsection{Inflaton decay}
In inflation scenarios where the inflaton is responsible for the density perturbation of the present universe, the inflaton should decay mainly to SM particles.
It can also partially decay to axions we are considering now (see for example Ref.~\cite{Dev:2013yza} for producing dark matter from the decay of inflaton).
In the sudden decay approximation, the axion abundance from such a partial decay of the inflaton is given by
\beq \label{Ya-inflaton}
Y_a = \frac{3}{4} \frac{T_{\rm R}}{m_I} \frac{{\rm Br}_{I \to aa}}{{\rm Br}_{I \to \rm SM}} \simeq 0.75 \,{\rm Br}_{I \to aa} \l( \frac{T_{\rm R}}{m_I} \r) 
\eeq
where ${\rm Br}_{I \to aa}$ and ${\rm Br}_{I \to \rm SM}$ are respectively the branching fractions of inflaton ($I$) to axions and to SM particles, and $T_{\rm R}$ and $m_I$ are the reheating temperature and mass of inflaton, respectively.
We assumed ${\rm Br}_{I \to \rm SM} \simeq 1$ in the far right side of \eq{Ya-inflaton}.
Comparing \eq{Ya-inflaton} to Eqs.~(\ref{TX-struc-bnd})
\footnote{We assumed that the momentum of inflaton was red-shifted and negligible relative to its mass when it decayed.
Otherwise, the dependence on the momentum of inflaton would have had to be included. We thank Anupam Mazumdar for pointing this out.}
and~(\ref{Ya-obs}), we find that a right amount of the axion relic density can be obtained while satisfying the constraint from structure formation, provided that 
\beq
{\rm Br}_{I \to aa} \lesssim 4.6 \times 10^{-4} \l( \frac{\lambda_{\rm fs}^{c}}{0.12 {\rm Mpc}} \r) \l( \frac{g_*(T_{\rm R})}{200} \r)^{1/4} \,.
\eeq

\subsection{Saxion decay}
The saxion, the radial component of the complex field containing the axion, can play a crucial role in the axion production, since it can decay into a pair of axions via the axion kinetic term, 
\beq \label{axion-K}
\mathcal{L} \supset \frac{1}{2} \frac{s}{f_a} \l( \partial a \r)^2.
\eeq
In this case, the decay rate of the saxion to a pair of axions is given by
\beq \label{s-to-aa}
\Gamma_{s \to aa} = \frac{1}{64 \pi} \frac{m_s^3}{f_a^2} 
\eeq

In the presence of an extra heavy charged fermion coupled with the saxion via a Yukawa coupling in the following form:
\beq
{\cal L} \supset -\lambda s {\bar f} f,
\eeq
there is an additional contribution to the saxion decay rate,
\beq
\Gamma_{s\rightarrow {\bar f}f} 
= \frac{\lambda^2}{8\pi}\, m_s \left(1-\frac{4m^2_f}{m^2_s}\right)^{3/2} \nonumber \\
=\frac{c^2_3 m^2_f m_s}{8\pi f^2_a} \left(1-\frac{4m^2_f}{m^2_s}\right)^{3/2} 
\eeq
where in the far right-hand side, the Yukawa interaction from the effective Lagrangian in Eq.~(\ref{Lag}) was used.  Then, for $\Gamma_{s\rightarrow aa}\ll \Gamma_{s\rightarrow {\bar f}f}$, the branching fraction of the saxion decaying to a pair of axions is given by
\beq
{\rm Br}_{s \to aa} \simeq \frac{m^2_s}{8c^2_3 m^2_f} >\frac{1}{2c^2_3}. 
\eeq
Thus, for $|c_3| = \mathcal{O}(1-10)$ which may be a natural expectation, we obtain ${\rm Br}_{s \to aa} = \mathcal{O}(10^{-2}-0.1)$ that may match to all the requirements in some region of parameter space, as discussed in the following sections.

In the early universe, the saxion might be at the broken phase with $H_I \gtrsim m_s$, but it could undergo a coherent oscillation as $H \lesssim m_s$.
Then, the saxion decay might be the main source of axion production, although structure formation constrains the branching fraction to axions, similarly to the case of inflaton decay.
In the following argument, for simplicity, we express the full decay width of the saxion as
\beq \label{s-width}
\Gamma_s = \Gamma_{s \to aa} / {\rm Br}_{s \to aa}
\eeq
and we will use the sudden decay approximation for saxion decay.

\begin{itemize}
\item{High $T_{\rm R}$}

If the saxion decays while the universe is dominated by radiation, the temperature of the universe right after the decay of the saxion  is given by
\beq \label{Ts}
T_s = \l( \frac{\pi^2}{90} g_*(T_s) \r)^{-1/4} \sqrt{ \Gamma_{s \to aa} M_\pl / {\rm Br}_{s \to aa}}\,,
\eeq
where \eq{s-width} was used in the right-hand side. 
We find that the constraint from structure formation (\eq{TX-struc-bnd}) is now translated to 
\beq \label{Br-HTR-fs-bnd}
{\rm Br}_{s \to aa} \lesssim {\rm Br}_{s \to aa}^{\rm HTR, fs} \equiv 3.98 \times 10^{-3} \l( \frac{\lambda_{\rm fs}^{c}}{0.12 {\rm Mpc}} \r)^2 \l( \frac{m_s}{10^{10} \GeV} \r) \l( \frac{f_{a,0}}{f_a} \r)^2 \l( \frac{m_a}{7 \keV} \r)^2
\eeq 
The abundance of axions produced from the decay of the saxion is given by
\beq \label{Ya-saxion}
Y_a = 2 {\rm Br}_{s \to aa} Y_s(t_s),
\eeq 
where $t_s$ is the time when saxion decays.
If the saxion is produced via coherent oscillation at $t_{s, \rm osc}$ and decays at $t_s$ while the universe is dominated by radiation, we obtain $Y_s(t_s) = Y_s(t_{s, \rm osc})$ where
$Y_s(t_{s, \rm osc})$ is the initial abundance of the saxion in coherent oscillation, given by
\beq \label{Ys-osc}
Y_s(t_{s, \rm osc}) \sim \l( \frac{f_a}{M_\pl} \r)^2 \l( \frac{M_\pl}{m_s}\r)^{1/2} \simeq 4.3 \times 10^{-4} \l( \frac{f_a}{f_{a,0}} \r)^2 \l( \frac{10^{10} \GeV}{m_s} \r)^{1/2}.
\eeq
Here, we assumed that the initial saxion misalignment is almost the same as the axion decay constant $f_a$.
Then, from \eqss{Ya-obs}{Ya-saxion}{Ys-osc}, one finds the condition for a right amount of the axion dark matter,
\beq \label{Br-vs-ms}
{\rm Br}_{s \to aa} \sim {\rm Br}_{s \to aa}^{\rm HTR, rd} \equiv 8.0 \times 10^{-2} \l( \frac{7 \keV}{m_a} \r) \l( \frac{f_{a,0}}{f_a} \r)^{2} \l( \frac{m_s}{10^{10} \GeV} \r)^{1/2}
\eeq
Comparing \eqs{Br-HTR-fs-bnd}{Br-vs-ms}, we find that there could exist a viable parameter space for $f_a = f_{a,0}$ if $\lambda_{\rm fs}^{c}$ is pushed up to the value over $0.54 \,{\rm Mpc}$, corresponding to about $1 \keV$ mass of thermal warm dark matter. 
However, such a large free-streaming length is in a strong tension with structure formation, being excluded by the most recent analysis of Lyman alpha forest data at the $9\sigma$ CL \cite{Viel:2013fqw}.

We now take a smaller $f_a$, for which the axion does not saturate the observed relic density DM but the observed flux of the $X$-ray line can be still obtained.
Even in this case,  since the abundance of the axion is proportional to $f_a^2$,  the photon flux caused by the decay of the axion does not depend on $f_a$.
Therefore, the required ${\rm Br}_{s \to aa}$ is given by
\beq \label{Br-HTR-355}
{\rm Br}_{s \to aa} \sim {\rm Br}_{s \to aa}^{\rm HTR, 3.5} \equiv 8.0 \times 10^{-2} \l( \frac{7 \keV}{m_a} \r) \l( \frac{m_s}{10^{10} \GeV} \r)^{1/2}
\eeq
which does not depend on $f_a$.
Note that, if the abundance of the axion is much smaller than the observed relic density of DM, the constraint on the free-streaming length of the axion is irrelevant as long as the axion DM is non-relativistic around the epoch of CMB decoupling.
Particularly, 
if the energy density of axion is about or less than $\mathcal{O}(10)$\% of active neutrino species, the axion can play the role of dark radiation and only the correct photon spectrum would require $v_a^2 \lesssim \mathcal{O}(0.1)$ with $v_a$ being the velocity of axion at present.
However, it turns out that the requirement does not generate any new stronger constraint on ${\rm Br}_{s \to aa}$.
The ${\rm Br}_{s \to aa}$ required to have a fractional axion DM, $r \equiv \Omega_a/\Omega_{\rm CDM}$ is 
\beq \label{Br-HTR-sub-rd}
{\rm Br}_{s \to aa} \sim r \times {\rm Br}_{s \to aa}^{\rm HTR, rd}.
\eeq
The energy contribution of axion dark radiation is given by
\beq
\frac{\rho_a}{\rho_\nu} 
= r \frac{\Omega_{\rm CDM}}{\Omega_{\rm r}} \frac{\rho_{\rm r}}{\rho_\nu}
= \l[ 3 + \frac{8}{7} \l( \frac{11}{4} \r)^{4/3} \r] \frac{\Omega_{\rm CDM}}{\Omega_{\rm r}} r
\eeq
and it is constrained to be $\rho_a/\rho_\nu \leq \Delta N_{\rm eff}^{\rm bnd} \equiv 0.26$ \cite{Ade:2013zuv}.
Hence one finds
\beq
r \leq \l[ 3 + \frac{8}{7} \l( \frac{11}{4} \r)^{4/3} \r]^{-1} \l( \frac{\Omega_{\rm r}}{\Omega_{\rm CDM}} \r) \Delta N_{\rm eff}^{\rm bnd}
= 1.16 \times 10^{-5} \l( \frac{\Delta N_{\rm eff}^{\rm bnd}}{0.26} \r)
\eeq
Therefore, in order to have a sizable axion dark radiation matching to observation, one needs
\beq \label{Br-HTR-DR}
{\rm Br}_{s \to aa} \sim {\rm Br}_{s \to aa}^{\rm HTR,DR} \equiv \l[ 3 + \frac{8}{7} \l( \frac{11}{4} \r)^{4/3} \r]^{-1} \l( \frac{\Omega_{\rm r}}{\Omega_{\rm CDM}} \r) \Delta N_{\rm eff}^{\rm bnd} {\rm Br}_{s \to aa}^{\rm HTR, rd}
\eeq
Equating \eqs{Br-HTR-355}{Br-HTR-DR}, we find that the amount of hinted dark radiation and the $3.5 \keV$ $X$-ray line can be explained simultaneously if 
\beq
\frac{f_a}{f_{a,0}} \simeq 3.4 \times 10^{-3} \l( \frac{\Delta N_{\rm eff}}{0.26} \r)^{1/2}
\eeq 
and ${\rm Br}_{s \to aa} = {\rm Br}_{s \to aa}^{\rm HTR, 3.5}$.

If the decay of the saxion is delayed, the saxion may start to dominate the universe when $H \sim H_{\rm SD}$ with 
\beq \label{HSD}
H_{\rm SD} \sim m_s \l( \frac{f_a}{M_\pl} \r)^4
\eeq
and $H_{\rm SD} > \Gamma_s$, in other words, 
\beq \label{Br-HSD}
{\rm Br}_{s\to aa} > \frac{1}{64 \pi} \l( \frac{m_s}{f_a} \r)^2 \l( \frac{M_\pl}{f_a} \r)^4.
\eeq
However, comparing to \eq{Br-HTR-fs-bnd}, we find that the allowed region of parameter space is opened only if 
\beq \label{HSD-ms-bnd}
m_s \lesssim 10^4 \GeV \l( \frac{\lambda_{\rm fs}^{c}}{0.12 {\rm Mpc}} \r)^2 \l( \frac{f_a}{f_{a,0}} \r)^4 \l( \frac{m_a}{7 \keV} \r)^2.
\eeq
In this case, one finds  
\beq \label{Br-HTR-fs-bnd2}
{\rm Br}_{s \to aa}^{\rm HTR, fs} \lesssim 4 \times 10^{-9} \l( \frac{\lambda_{\rm fs}^{c}}{0.12 {\rm Mpc}} \r)^4 \l( \frac{f_a}{f_{a,0}} \r)^2 \l( \frac{m_a}{7 \keV} \r)^4\,.
\eeq

Now the abundance of the axion from the decay of the saxion is given by 
\bea \label{Ya-SD}
Y_a 
&\simeq& \frac{3}{2} {\rm Br}_{s \to aa} \,\frac{T_s}{m_s}  
\nonumber \\
&\simeq& 7.6 \times 10^{-5} \times \l(  \frac{{\rm Br}_{s \to aa}}{10^{-2}} \r)^{1/2}  \l( \frac{m_s}{10^6 \GeV} \r)^{1/2} \l( \frac{f_{a,0}}{f_a} \r)
\eea
where we assumed that the saxion decays mainly to SM particles, and used \eq{Ts} in the second line.
A right amount of the flux for the $3.5 \keV$ $X$-ray line is obtained if
\beq
\frac{Y_a}{f_a^2} = \frac{6.9 \times 10^{-5} \l( 7 \keV/m_a \r)}{f_{a,0}^2}.
\eeq
Combined with \eq{Ya-SD}, the above equation leads to 
\beq
{\rm Br}_{s \to aa}  = {\rm Br}_{s \to aa}^{\rm SD, 3.5} \equiv 13.3 \l( \frac{g_*(T_s)}{200} \r)^{1/2} \l( \frac{10^4 \GeV}{m_s} \r) \l( \frac{f_a}{f_{a,0}} \r)^6 \l( \frac{7 \keV}{m_a} \r)^2\,.
\eeq
Using \eq{HSD-ms-bnd}, one finds 
\beq
{\rm Br}_{s \to aa}^{\rm SD, 3.5} > 13.3 \l( \frac{g_*(T_s)}{200} \r)^{1/2} \l( \frac{0.12 {\rm Mpc}}{\lambda_{\rm fs}^{c}} \r)^2 \l( \frac{f_a}{f_{a,0}} \r)^2 \l( \frac{7 \keV}{m_a} \r)^4,
\eeq
which contradicts \eq{Br-HTR-fs-bnd2}.
Therefore, it is not possible to get a right amount of photon flux while satisfying the constraint from structure formation in this case of saxion domination.

\item{Low $T_{\rm R}$}

Inflaton decay might be delayed to a time after axion production, i.e., $\Gamma_I < \Gamma_s$ which requires
\beq \label{Br-LTR-bnd}
{\rm Br}_{s \to aa} < {\rm Br}_{s \to aa}^{\rm LTR} \equiv \Gamma_{s \to aa}/\Gamma_I = \l( \frac{\pi^2}{90} g_*(T_{\rm R}) \r)^{-1/2} \frac{1}{64 \pi} \frac{m_s}{M_\pl} \l( \frac{M_\pl}{f_a} \r)^2 \l( \frac{m_s}{T_{\rm R}} \r)^2.
\eeq
In this case, the free-streaming length is given by
\bea
\lambda_{\rm fs}
&\approx& \frac{1}{H_0} \l( \frac{H_0}{\Gamma_s} \r)^{2/3} \l( \frac{\Gamma_I}{H_0} \r)^{1/6} \l( \frac{m_X/2}{m_a} \r) \l( \frac{T_{\rm eq}}{T_0} \r)^{1/4}
\nonumber \\
&& \times \l\{ 1 + \frac{1}{2} \ln \l[ \l( \frac{\Gamma_s}{H_0} \r)^{4/3} \l( \frac{H_0}{\Gamma_I} \r)^{1/3} \l( \frac{m_a}{m_X/2} \r)^2 \l( \frac{T_0}{T_{\rm eq}} \r)^{3/2} \r] \r\}
\eea
and the constraint from structure formation (\eq{lfs-bnd}) is interpreted as
\bea \label{Br-LTR-struc-bnd}
{\rm Br}_{s \to aa} &\lesssim& {\rm Br}_{s \to aa}^{\rm LTR,fs} 
\nonumber \\
&\equiv& \l( \frac{\lambda_{\rm fs}^{c}}{0.12 {\rm Mpc}} \r)^{3/2} \l( \frac{H_0}{5.33 \times 10^{-38} \GeV} \r)^{3/2} \l( \frac{\Gamma_I}{H_0} \r)^{1/2} \l( \frac{m_a}{m_s/2} \r)^{3/2} \l( \frac{T_0}{T_{\rm eq}} \r)^{3/8} 
\nonumber \\
&& \times \l\{ 1 + \frac{1}{2} \ln \l[ \l( \frac{\Gamma_s}{H_0} \r)^{4/3} \l( \frac{H_0}{\Gamma_I} \r)^{1/3} \l( \frac{m_a}{m_X/2} \r)^2 \l( \frac{T_0}{T_{\rm eq}} \r)^{3/2} \r] \r\}^{-3/2} {\rm Br}_{s \to aa}^{LTR}
\nonumber \\
&=& 1.5 \times 10^{-12} \l( \frac{\lambda_{\rm fs}^{c}}{0.12 {\rm Mpc}} \r)^{3/2} \l( \frac{m_a}{7 \keV} \r)^{3/2} \l( \frac{f_{a,0}}{f_a} \r) \l( {\rm Br}_{s \to aa}^{\rm LTR} \r)^{1/2}
\eea
 in the last line we used an approximation, $\l\{ \cdots \r\} = 8.26$.
If the saxion starts its coherent oscillation as $H \lesssim m_s$, for $f_a$ around of larger than intermediate scale which may appear naturally in theories beyond the standard model (for example, SUSY or string theories), the abundance of the axion when the inflaton decays is given as
\bea
Y_a(T_{\rm R}) 
&\simeq& \frac{1}{2} {\rm Br}_{s \to aa} \l( \frac{f_a}{M_\pl} \r)^2 \l( \frac{T_{\rm R}}{m_s}\r)
\nonumber \\
&\lesssim& 2.5 \times 10^{-22} \l( \frac{m_a}{7 \keV} \r)^{3/2} \l( \frac{m_s}{10^{10} \GeV} \r)^{1/2}
\eea
where in the last line we used \eq{Br-LTR-struc-bnd}.
We find that if the inflaton decayed after axion production, the axion abundance coming from the decay of the saxion would have turned out to be too small.

\end{itemize}

\subsection{Axion misalignment}
The $\keV$-scale mass of the axion is far below the typical expansion rate of inflation.
In addition, the mass of the axion is generated by the anomaly only after the associated symmetry is broken.
Hence, if the symmetry were broken after inflation, a typical axion misalignment would have been of order of the axion decay constant.
On the other hand, if the symmetry were broken before or during inflation, the amount of misalignment could have been much smaller than the axion decay constant. 
In the following argument, we assume the latter case to allow a wide range of misaligned axion field values.

\begin{itemize}
\item{High $T_{\rm R}$}

The energy density of the axion at the onset of oscillation can be expressed as
\beq \label{rho-axion}
\rho_{a,\rm osc} = \frac{1}{2} m_a^2 \theta_{\rm osc}^2 f_a^2
\eeq
where $\theta_{\rm osc}$ is the initial misalignment angle, and $\theta_{\rm osc} \ll 1$ is assumed.
We assume that the inflaton decayed before the axion started its oscillation.
Then, the present abundance of the misaligned axion is 
\beq
Y_a = \frac{\sqrt{3}}{8} \l( \frac{\pi^2}{90} g_*(T_{\rm osc}) \r)^{-1/4} \theta_{\rm osc}^2 \l( \frac{f_a}{M_\pl} \r)^2 \l( \frac{M_\pl}{m_a} \r)^{1/2}
\eeq 
This can be consistent with the observed relic density if
\beq
\theta_{\rm osc} \lesssim 2 \times 10^{-4} \l( \frac{f_{a,0}}{f_a} \r) \l( \frac{7 \keV}{m_a} \r)^{1/2}
\eeq
where we used $g_*(T_{\rm osc})=200$, and the upper-bound saturates the relic density.

A remark is in order here. 
Considering the primordial inflation, one notice that $\theta_{\rm osc} \ll1$ requires that the modulus associated with the axion should be in the broken phase during inflation so as for a Hubble patch to be occupied by a particular value $\theta_{\rm osc}$. 
In addition, as already shown in \eq{HI-bnd}, for $\theta_{\rm osc} \lesssim 10^{-4}$, the expansion rate of the primordial inflation ($H_I$) should be less than of order of $\mathcal{O}(10^6) \GeV$ in order not to produce too much iso-curvature perturbation caused by perturbations of $\theta_{\rm osc}$.

The recent data of BICEP2 hinted that the Hubble scale during inflation is $H_I \sim 10^{14} \GeV$ \cite{bicep2}.
If it is confirmed, axion cannot be the main component of DM, unless $f_a \gtrsim M_\pl$.
However, it may be still possible to obtain a right amount of the $X$-ray line flux for $f_a \ll f_{a,0}$ while satisfying the constraint from iso-curvature perturbation.
For example, the photon flux from the decay of the misaligned axion is proportional to $\theta_{\rm osc}^2$ and does not depend on $f_a$ as long as the initial abundance of the axion is given by \eq{rho-axion}.
Since $r$ in \eq{HI-bnd} is proportional to $f_a^2$, one can take $f_a \ll f_{a,0}$ to push up the bound of $H_I$ to $\mathcal{O}(10^{14}) \GeV$ hinted by BICEP2 data, while keeping $\theta_{\rm osc} \sim 2 \times 10^{-4}$.

\item{Low $T_{\rm R}$}

The inflaton might decay at a very late time, and the axion might start its oscillation when the universe was still dominated by the inflaton.
In this case, the axion abundance is given by
\beq
Y_a = \frac{1}{8} \frac{T_{\rm R}}{m_a} \theta_{\rm osc}^2 \l( \frac{f_a}{M_\pl} \r)^2 
\eeq
where $T_{\rm R}$ is the reheating temperature of the inflaton.
Hence, compared to \eq{Ya-obs}, $\theta_{\rm osc}$ is upper-bounded as
\beq
\theta_{\rm osc} \lesssim 0.4 \l( \frac{f_{a,0}}{f_a} \r) \l( \frac{1 \GeV}{T_{\rm R}} \r)^{1/2} 
\eeq
Note that, since $T_{\rm R} \gtrsim 10 \MeV$, $\theta_{\rm osc} \sim \mathcal{O}(0.1-1)$ is allowed and the saxion can be in the symmetric phase during inflation although we then may have to worry about the domain wall problem.
The iso-curvature perturbation bound on $H_I$ in eq. (\ref{HI-bnd}) may be satisfied marginally, being compatible with BICEP2 data, within some errors, for $\theta_{\rm osc} \sim \mathcal{O}(1)$ and $f_a^I \sim \mathcal{O}(10^{16-17}) \GeV$.

\end{itemize}

\section{Conclusion}
\label{sec:conclusion}

We proposed a simple model for the $\keV$-scale axion dark matter whose decay product into monochromatic photons can be the source of the recently reported $X$-ray spectrum at about $3.5 \keV$.
Such a light axion can be produced from the decay of a heavy particle or from the coherent oscillation of the axion caused by misalignment.
We showed how the keV-sale axion model is constrained by structure formation and iso-curvature perturbation, depending on the axion production mechanisms and the amount of dark matter relic density.

We found that axions produced from the inflaton decay can saturate the observed relic density of dark matter and satisfy the constraint from structure formation, provided that the reheating temperature was not smaller than the inflaton mass by an order of magnitude and the branching fraction of the inflaton decay to axions was less than about $\mathcal{O}(10^{-4})$.
In the case of saxion decay, on the other hand, if the saxion was coherently produced in a broken phase and decayed during the epoch of radiation domination after inflaton decay, the axions produced from the saxion decay would be in a strong tension with the bounds from the Lyman-alpha forest data. 
The tension can be removed if axions produced from the saxion decay is of a subdominant contribution of the dark matter relic density at present.
This, in turn, requires a smaller axion decay constant to keep the observed $3.5$ keV $X$-ray line the same. In this case, it is interesting to notice that the axions can play the role of dark radiation simultaneously, being compatible with Planck data.
If the saxion of coherent oscillation dominated the universe at a later time, the constraint from structure formation would have made it difficult to accommodate the $3.5 \keV$ $X$-ray line signal consistently.
Also, if the inflaton decayed at a late time after the decay of the saxion, the axion relic density could have not saturated the observed dark matter relic density, without disturbing the structure formation due to relativistic axions.

We also discussed the axion misalignment for the axion production. In this case, if the inflaton decayed before the axion started its oscillation, the misalignment angle $\theta_{\rm osc}$ should be about $10^{-4}$ to saturate the relic density.
On the other hand, if the reheating temperature of primordial inflation is about $10 \MeV$, it is possible to have a natural value of $\theta_{\rm osc} \sim 0.1$.
In the case of saxion domination, the parameter space of the misaligned axion is more constrained, due to the axions produced from the saxion decay.

\section*{Acknowledgements}
This research is supported in part by Basic Science Research Program through the National Research Foundation of Korea funded by the Ministry of Education, Science and Technology (2013R1A1A2007919) (HML),  
(2011-0010294), (2011-0029758) and (2013R1A1A2064120) (SCP) and (2012R1A2A1A01006053) (WIP). HML is also supported by the Chung-Ang University Research Grants in 2014.


\begin{thebibliography}{999}



\bibitem{Xray1}
  E.~Bulbul, M.~Markevitch, A.~Foster, R.~K.~Smith, M.~Loewenstein and S.~W.~Randall,
  arXiv:1402.2301 [astro-ph.CO].



\bibitem{Xray2}
  A.~Boyarsky, O.~Ruchayskiy, D.~Iakubovskyi and J.~Franse,
  arXiv:1402.4119 [astro-ph.CO].




\bibitem{astroH}
  T.~Takahashi, K.~Mitsuda, R.~Kelley, H.~Aharonian, F.~Aarts, H.~Akamatsu, F.~Akimoto and S.~Allen {\it et al.},
  arXiv:1210.4378 [astro-ph.IM].
  

\bibitem{takahashi1}
  H.~Ishida, K.~S.~Jeong and F.~Takahashi,
  arXiv:1402.5837 [hep-ph].


\bibitem{finkbeiner}
  D.~P.~Finkbeiner and N.~Weiner,
  arXiv:1402.6671 [hep-ph];
  J.~-C.~Park, S.~C.~Park and K.~Kong,
  Phys.\ Lett.\ B {\bf 733}, 217 (2014)
  [arXiv:1403.1536 [hep-ph]].


\bibitem{axionreview}
  J.~E.~Kim and G.~Carosi,
  Rev.\ Mod.\ Phys.\  {\bf 82} (2010) 557
  [arXiv:0807.3125 [hep-ph]].



\bibitem{takahashi2}
  T.~Higaki, K.~S.~Jeong and F.~Takahashi,
  arXiv:1402.6965 [hep-ph].

\bibitem{ringwald} 
  J.~Jaeckel, J.~Redondo and A.~Ringwald,
  arXiv:1402.7335 [hep-ph].


\bibitem{susyaxion}
  J.~E.~Kim and H.~P.~Nilles,
  Phys.\ Lett.\ B {\bf 138} (1984) 150;
  T.~Higaki and R.~Kitano,
  Phys.\ Rev.\ D {\bf 86} (2012) 075027
  [arXiv:1104.0170 [hep-ph]].


\bibitem{WDMconstraint1} 
  M.~Viel, J.~Lesgourgues, M.~G.~Haehnelt, S.~Matarrese and A.~Riotto,
  Phys.\ Rev.\ D {\bf 71}, 063534 (2005)
  [astro-ph/0501562].

\bibitem{WDMconstraint2} 
  R.~S.~de Souza, A.~Mesinger, A.~Ferrara, Z.~Haiman, R.~Perna and N.~Yoshida,
  Mon.\  Not.\  Roy.\  Astron.\  Soc.\  {\bf 432}, 3218 (2013)
  [arXiv:1303.5060 [astro-ph.CO]].

\bibitem{Viel:2013fqw} 
  M.~Viel, G.~D.~Becker, J.~S.~Bolton and M.~G.~Haehnelt,
  Phys.\ Rev.\ D {\bf 88}, no. 4, 043502 (2013)
  [arXiv:1306.2314 [astro-ph.CO]].

\bibitem{lyalpha}
  K.~Markovic and M.~Viel,
  Publications of the Astronomical Society of Australia / Volume 31
  / January 2014, e006 (20 pages)
  [arXiv:1311.5223 [astro-ph.CO]].


\bibitem{dSph}
  A.~Boyarsky, O.~Ruchayskiy and D.~Iakubovskyi,
  JCAP {\bf 0903} (2009) 005
  [arXiv:0808.3902 [hep-ph]];
  D.~Gorbunov, A.~Khmelnitsky and V.~Rubakov,
  JCAP {\bf 0810} (2008) 041
  [arXiv:0808.3910 [hep-ph]];
  S.~Horiuchi, P.~J.~Humphrey, J.~Onorbe, K.~N.~Abazajian, M.~Kaplinghat and S.~Garrison-Kimmel,
  Phys.\ Rev.\ D {\bf 89} (2014) 025017
  [arXiv:1311.0282 [astro-ph.CO]].

\bibitem{Planckdata} 
  P.~A.~R.~Ade {\it et al.}  [Planck Collaboration],
  arXiv:1303.5082 [astro-ph.CO].


\bibitem{axionisocurvature} 
  T.~Kobayashi, R.~Kurematsu and F.~Takahashi,
  JCAP {\bf 1309}, 032 (2013)
  [arXiv:1304.0922 [hep-ph]].


\bibitem{Dev:2013yza} 
  P.~S.~Bhupal Dev, A.~Mazumdar and S.~Qutub,
  arXiv:1311.5297 [hep-ph].

\bibitem{Ade:2013zuv} 
  P.~A.~R.~Ade {\it et al.}  [Planck Collaboration],
  arXiv:1303.5076 [astro-ph.CO].

\bibitem{bicep2}
  P.~A.~R.~Ade {\it et al.}  [BICEP2 Collaboration],
  arXiv:1403.3985 [astro-ph.CO].



\end{thebibliography}
\end{document}